\documentclass[useAMS,usenatbib ]{mn2e}
\usepackage{graphicx}
\usepackage{epstopdf}

\title[Galaxy Wings and Residual Fluctuations]{The Contribution of Faint Galaxy Wings to Source-subtracted Near-infrared Background Fluctuations}
\author[R.L. Donnerstein]{R.L. Donnerstein$^{1}$\thanks{E-mail:
donnerst@email.arizona.edu} \\
$^{1}$The University of Arizona, Tucson, AZ 85724, USA}
\begin{document}

\date{Accepted 2015 February 16.  Received 2015 February 5; in original form 2014 July 11}

\pagerange{\pageref{firstpage}--\pageref{lastpage}} \pubyear{2002}

\maketitle

\label{firstpage}

\begin{abstract}
The source-subtracted, 1.1 and 1.6 $\mu$m NICMOS images used in earlier analyses of the near-infrared Hubble Ultra Deep Field contained residual flux in extended wings of identified sources that contributed an unknown amount to fluctuation power. When compared to the original results, a reanalysis after subtracting this residual flux shows that mean-square and rms fluctuations decrease a maximum of 52 and 31 per cent at 1.6 $\mu$m and 50 and 30 per cent at 1.1 $\mu$m. However, total mean-square fluctuations above 0.5 arcsec only decrease 6.5 and 1.4 per cent at 1.6 and 1.1 $\mu$m, respectively.  These changes would not affect any published conclusions based on the prior analyses. These results exclude previous suggestions that extended wings of detected galaxies may be a major contributor to the source-subtracted near-infrared background and confirm that most fluctuation power in these images must be explained by other means. 
\end{abstract}

\begin{keywords}
infrared: diffuse background -- methods: data analysis -- cosmology: observations.
\end{keywords}

\section{Introduction} \label{Introduction}
\begin{figure}
\center{\includegraphics[ width =\columnwidth]{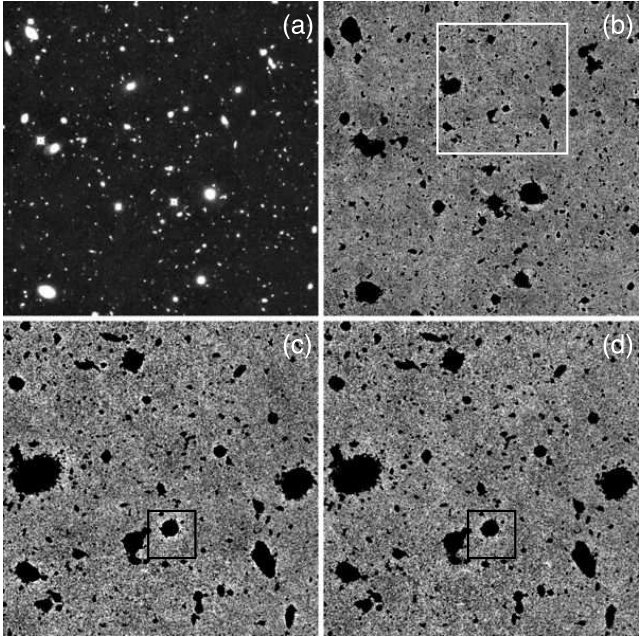}}
\caption{Image examples. (a): Original, full F160W image. (b): Source-subtracted, full F160W image. White box outlines enlarged region shown in the lower panels and in Fig.~\ref{Im-Exp_x3}. (c): Enlarged F160W image. Residual flux is evident in source wings. Black box outlines the source used for the detailed example shown in Fig.~\ref{Im-ExpExample}. (d): Enlarged F110W image. Source wings are still seen, but are not as prominent as in the F160W image. Panels (b)--(d) have the same linear stretch. Unless otherwise noted, masked pixels in figures are set to black for illustration purposes. These pixels are assigned values of zero for fluctuation calculations.}
 \label{Im-fields}
\end{figure}

Multiple theories have been offered to explain residual fluctuations in the near-infrared background (NIRB). After analysing images from \textit{Spitzer}/IRAC \citep{k3,k5} and \textit{AKARI}/IRC \citep{m3}, the authors concluded that NIRB fluctuations are unlikely to result from known galaxy populations and are consistent with the earliest star formation era. On the other hand, evaluation of NICMOS images from the Hubble Ultra Deep Field (NUDF) suggested that source-subtracted fluctuations result from normal galaxies at $z<$ 8, with most power in the redshift range of 0.5--1.5 \citep{t1,t4}. More recently, other explanations have been proposed to account for the majority of near-infrared fluctuations, including intrahalo light (IHL) from dark-matter haloes \citep{c4,z1} and early, direct collapse black holes (DCBHs) as suggested by \citet{y2}. Conclusions are frequently based, in part, by the observation that NIRB fluctuations measured by different instruments over a broad range of wavelengths are consistently much higher than those predicted for known populations \citep{ h2} or high-redshift galaxies \citep{c3,y1,f1}. Fluctuation measurements considered in these studies include those derived from 1.6 $\mu$m NUDF observations as described by \cite{t1}. All of these interpretations depend upon the amplitudes and shapes of observed fluctuations, which, in turn, are related to the completeness of the source-extraction process. \citet{h2} appropriately noted that the source-subtracted NUDF images analysed by \cite{t1} included residual flux in the extended outer edges of galaxies (wings). This residual flux would impact fluctuation estimates to an unknown extent and could potentially affect conclusions derived from their spectra. As described by \citet{t2}, source subtraction for these images was deliberatively conservative to minimise the possibility of false sources appearing in their catalogue. This resulted in incomplete subtraction of detected objects with visible flux remaining in the wings of identified sources (Fig.~\ref{Im-fields}c). If these unsubtracted wings are a substantial component of measured fluctuations, conclusions drawn from their results might have to be reconsidered. While \citet{k3} and \citet{a1} showed that residual wings contributed only small amounts to the power spectra of \textit{Spitzer}/IRAC images, their methodology was limited by the resolution of this system. The higher resolution of \textit{Hubble}/NICMOS observations allows a more robust approach for assessing these effects. This paper reanalyses the NUDF images used by \cite{t1} to quantify the contributions of these unsubtracted, extended galaxy wings to measured fluctuations.  
\section{Observations and Data Reduction}
All images and tables used in this study are available online.\footnote{http://archive.stsci.edu/pub/hlsp/udf/nicmos-treasury/} Detailed descriptions of the processing of NUDF images are documented in \citet{t3}. Only overviews of portions relevant to this study are presented. 
\subsection{Image acquisition and processing}
NUDF observations cover a $144 \times 144$ arcsec area and are obtained with two filters centred on wavelengths of 1.1 (F110W) and 1.6 (F160W) $\mu$m. Field position on the camera is dithered by more than the average source size during separate acquisitions which results in a median image dominated by zodiacal emission \citep{t1}. Subtracting this median image from the original images effectively removes smooth background components, including the zodiacal light. Images are then combined in a drizzle procedure such that the original 0.2 arcsec pixels are converted to 0.09 arcsec to match a $3 \times 3$ binning of the 0.03 arcsec \textit{Hubble} Advanced Camera for Surveys (ACS) images.
\subsection{Source extraction}
The source extraction procedure is described in \citet{t2} and is briefly summarised here. Using a process described by \citet{s1}, all four ACS and the two NICMOS bands are used to identify those pixels with a sufficient signal to noise ratio to be considered part of a real source. Source extraction is then done with SExtractor (SE) in the two-image mode as described by \citet{b1}. This process uniquely identifies individual sources and their locations on the image.\footnote{http://archive.stsci.edu/pub/hlsp/udf/nicmos-treasury/table/szalayf160wnfdet.fit} The ACS optical images have much longer integration times than NICMOS images and, therefore, identify sources not apparent in the NUDF images alone. This results in more complete source detection than would be possible using only NICMOS observations. A total of 4702 objects are listed in the NUDF catalogue generated by SE\footnote{http://archive.stsci.edu/pub/hlsp/udf/nicmos-treasury/table/newudfcommatable.txt}, 4276 of which remain after trimming. Sources identified by this method are then subtracted (masked) from the image by setting their corresponding pixels to zero. As shown in Fig.~\ref{Im-fields}, this process may fail to subtract outer portions of galaxies more prominent in the NICMOS images. Detector quantum efficiency is less at 1.1 $\mu$m than at 1.6 $\mu$m causing signals from the outer regions of galaxies to blend more quickly into background noise. This makes source subtraction for the F110W image in Fig.~\ref{Im-fields}d to falsely appear more complete than that for the F160W image (Fig.~\ref{Im-fields}c). 
\subsection{Fluctuation analysis}
Fluctuation analyses are performed as described in detail in Appendix A of \citet{t1} and are similar to those used by others \citep{k3,c2}. The original image in units of ADU s$^{-1}$ is converted to Jy pixel$^{-1}$ using standard procedures for NICMOS images. Pixel values are then converted to W m$^{-2}$ sr$^{-1}$ by multiplying by the corresponding frequency for 1.1 or 1.6 $\mu$m and dividing by the solid angle subtended by each pixel. Each source-subtracted image is then multiplied by a normalised weighting image to account for different pixel observing times. Prior to Fourier transformation, the mean value of the unmasked pixels are subtracted from this result to produce an image with a mean of zero. The two-dimensional Fourier transform of the resultant image, $f(\bmath{q})$, is computed using an FFT, where $\bmath{q}$ is the wave vector. The value of the wavenumber, $q$, and the amplitude of the transform, $|f(q)|$, is calculated for each point in the returned array. A log-spaced vector of $q$ values is generated and the average value of $|f(q)|^2$, $P_2$, is computed for each of the bins defined by this vector. Mean-square fluctuations, $\mathcal{F}^2(\theta)\approx q^2P_2/(2\pi)$, is then calculated, where $\theta$ is $2\pi/q$ converted into arcsec. Mean-square fluctuations are divided by the fraction of unsubtracted pixels in the image to compensate for area lost by masking \citep{a1}. Unless otherwise noted, error bars shown in figures represent one standard deviation of the Poisson errors, $P_2/\sqrt{N_q}$, where ${N_q}$ is the number of Fourier elements in the associated bin. Because fluctuations may be reported as the mean-square \citep{k5,f1} or root-mean-square (rms), $\sqrt{q^2P_2/(2\pi)}$ \citep{k3,t1,h2}, both of these are calculated when appropriate. 
\begin{figure*}
 \center{\includegraphics[ width =\textwidth ,trim = 0in 0in 0in 0in, clip]{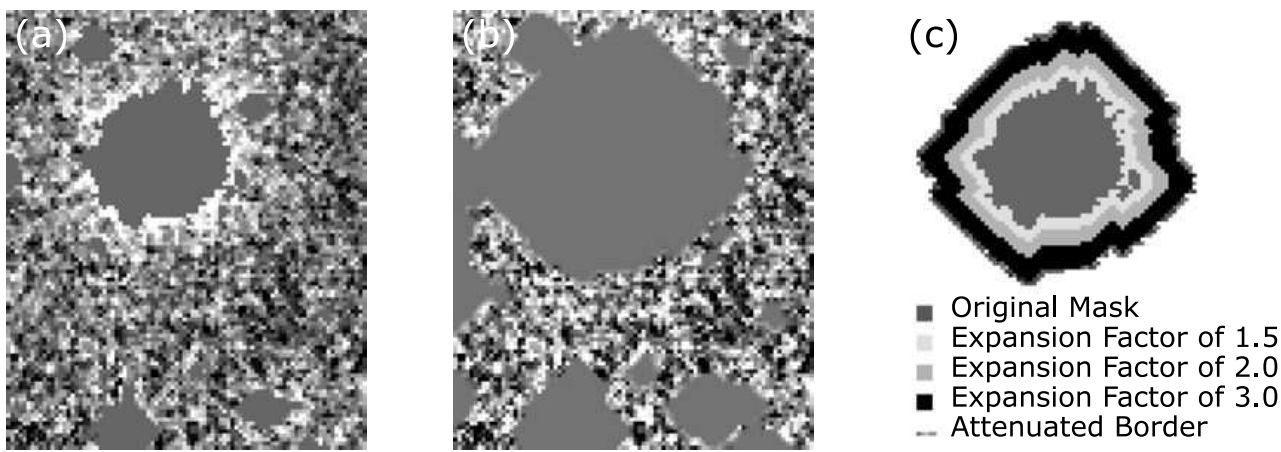}}
 \caption{Example of mask expansion of the source outlined in Fig~\ref{Im-fields}. (a): Original source subtraction showing significant residual wings. (b): The mask area is enlarged by a factor of three. (c): Source masks for expansions of 1.5, 2.0 and 3.0. Masked pixels are assigned values of zero in Panels (a) and (b). }
 \label{Im-ExpExample}
\end{figure*}
 \begin{figure}
 \center{\includegraphics[ width = \columnwidth,trim = 0in 0in 0in 0in, clip]{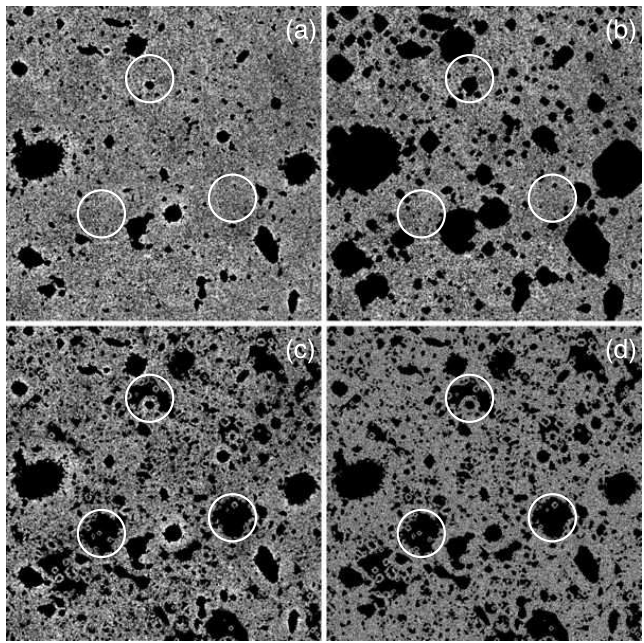}}
 \caption{F160W images used to estimate effects of expanding individual source mask areas. (a): Original, source-subtracted image before mask expansion. (b): Mask-expanded image with source mask areas expanded by a factor of three. When compared to Panel (a), residual circumferential flux is markedly reduced. (c): Control Image. Random source masks are placed in random locations to exactly match subtracted pixel count of the mask-expanded image shown in Panel (b). (d): Noise image. Unsubtracted (non-zero) pixels in the control image are replaced with Gaussian noise. Circles outline three of many randomly placed source masks generated for control images (lower panels) that are not present in the original or mask-expanded images (upper panels). The upper circle shows that wings of identified sources are not masked in the control images. }
 \label{Im-Exp_x3}
\end{figure}

\section{Flux from extended wings}
The contribution of unsubtracted flux in galaxy wings to total fluctuations is estimated by incrementally expanding subtracted regions (masks) for individual sources. While this expansion effectively eliminates residual flux in wings, the masking process itself can distort the Fourier power spectra. To account for this distortion, comparisons are made to control images with the same number of unsubtracted pixels (see Section~\ref{ss-Control}). 
\subsection{Expansion of individual source masks}\label{ss-Expanded}
To quantify contributions from galaxy wings over a range of brightness, all source masks are uniformly expanded in a predefined manner. This is possible because the extraction process identifies specific objects, thereby allowing mask expansion tailored to individual source size. Source mask expansions are evaluated over a range of 1.1 to 6.0 times the original source-mask pixel count. Fig.~\ref{Im-ExpExample} demonstrates this process applied to the single galaxy outlined in the bottom panels of Fig.~\ref{Im-fields}. In the left panel, residual flux is seen extending beyond the original subtracted region. The middle panel shows that residual flux from the same object is markedly reduced after the mask area is tripled. The right panel demonstrates the expansion process in more detail for expansion factors of 1.5, 2.0 and 3.0. Individual source masks are circumferentially extended in one pixel increments until the number of masked pixels is the desired multiple of its original value. This process will significantly over-subtract small objects where one pixel represents more than 10 per cent of the desired mask area.  This problem is minimised by expanding the outermost portion of the mask on an image resampled with pixels 1/10 their original edge length using a cubic spline interpolation. These outermost pixels are then rebinned back to their original size by averaging over the 100 resampled pixels.  As shown in the right panel of Fig.~\ref{Im-ExpExample}, this results in an attenuation of these outer pixels rather than them being set to zero. The effect of this attenuation is included when calculating the total number of masked pixels on an image.  The equivalent percentage of image pixels masked for a range of expansion factors is shown in Table \ref{tbl-maskfrac}. 

\begin{figure*}
 \center{\includegraphics[width = \textwidth,trim = 0in 0in 0in 0in, clip]{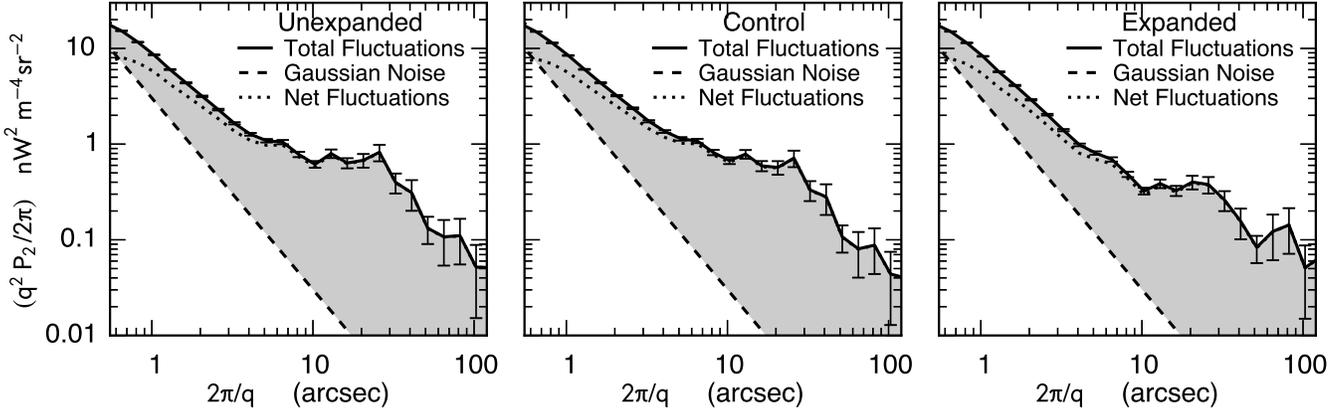}}
 \caption{Mean-square fluctuations for unexpanded, control, and mask-expanded F160W images after expanding source-mask areas by a factor of four. Gaussian noise fluctuations are estimated separately for each image. Error bars represent 1$\sigma$ of Poisson errors. Shaded areas correspond to fluctuations above noise.} 
\label{Plt-Exp_x3_Example}
\end{figure*}
\begin{table}
\begin{minipage}{\columnwidth}
 \caption{Percentage of masked image pixels for selected expansions.  Values include effects of attenuated pixels (see Section \ref{ss-Expanded}).}
 \label{tbl-maskfrac}
 \begin{tabular}{lc}
 \hline
 Expansion factor&Per cent of masked pixels (\%)\\
 \hline
1.0 (baseline)& 13\\
2.0 & 22\\
3.0 & 30\\
4.0 & 38\\
5.0 & 44\\
6.0 & 50\\
\hline
\end{tabular}
\end{minipage}
\end{table}

An example of an F160W image where source mask sizes are expanded by a factor of three is shown in Fig.~\ref{Im-Exp_x3}b. When compared to the baseline source-subtracted image shown in Fig.~\ref{Im-Exp_x3}a, it is apparent that much of the residual flux in wings is eliminated.

\subsection{Generation of control images}\label{ss-Control}
The masking process itself can cause a distortion of fluctuation spectra that must be considered when making comparisons \citep{a1,k5}. Images are masked by multiplying the image field by a separate mask with pixel values of one everywhere except for source-subtracted regions that have pixel values of zero or, as shown in Fig. 2, a transmission factor. The Fourier transform of this multiplication is equivalent to the convolution of the Fourier transform of the mask with that of the unmasked image. Spectral properties of the mask are influenced by both the locations and sizes of individual source masks. The larger, brighter sources are not evenly distributed throughout the NUDF field and tend to be grouped towards the left side of the image (Fig. 1). Masking this uneven population of bright sources would be expected to generate some redistribution of fluctuation power that is not representative of the true background spectrum. Because of these effects, mask-expanded spectra are compared to those of control images generated by adding randomly placed source masks to the original source-subtracted NUDF image. This additional masking is done by randomly selecting individual source identification numbers from among the 4276 objects identified in the source extraction process. The mask from the selected object is then placed in a random location on the image. This process continues until the total number of masked pixels in the control image exactly matches that of its associated mask-expanded image. During this process it is important that the randomly placed source-masks remain on background pixels and do not subtract galaxy wings.  This is accomplished by not allowing a randomly placed mask to encroach on a region encompassing four times the original mask size of an identified source. A region this size encloses essentially all of the residual wing flux associated with an object (see Section~\ref{ss-results}).   An image created in this manner is shown in Fig.~\ref{Im-Exp_x3}c. Because locations of the randomly placed source-masks will affect fluctuation calculations, 1000 realisations of the control images are generated. Contributions of noise to fluctuation power are estimated by replacing all nonzero pixels of an image with Gaussian noise of the same mean and standard deviation as background sky (Fig.~\ref{Im-Exp_x3}d). This is done for all images. Averaged fluctuation spectra of the 1000 control images and respective noise images are used for further calculations. 

\subsection{Effects of mask expansion on fluctuations}\label{ss-Effects}

\begin{figure}
 \center{\includegraphics[width = \columnwidth,trim = 0in 0in 0in 0in, clip]{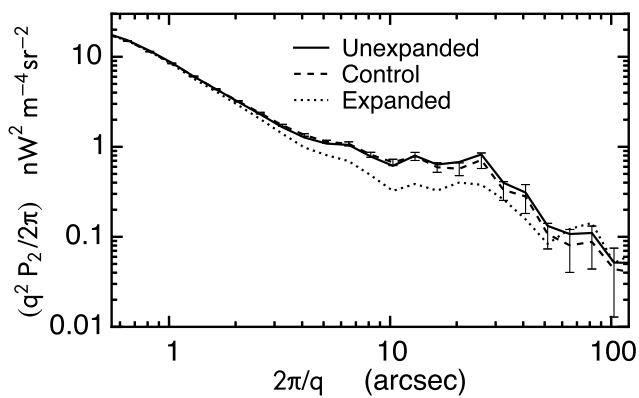}}
 \caption{Mean-square fluctuations for unexpanded, control, and mask-expanded F160W images after expanding source-mask areas by a factor of four. Error bars represent 1$\sigma$ of Poisson errors in the control image.}
 \label{Plt-Exp_x3_Overplots}
\end{figure}
 
 \begin{figure}
 \center{\includegraphics[width = \columnwidth,trim = 0in 0in 0in 0in, clip]{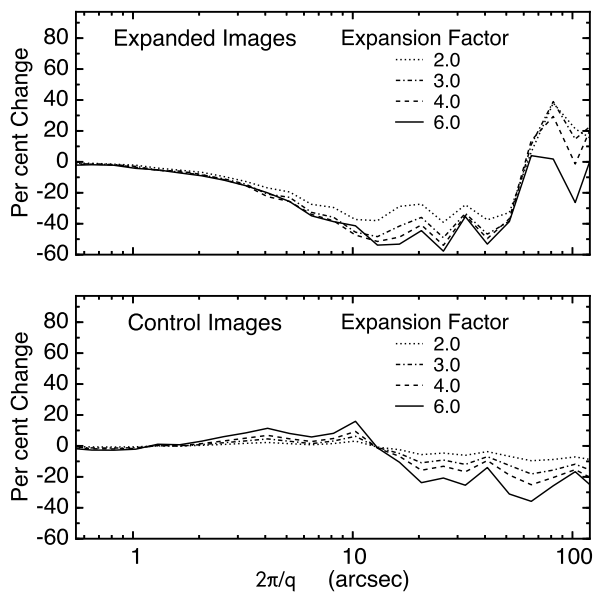}}
 \caption{Mean-square fluctuation changes at 1.6 $\mu$m relative to unexpanded, source-subtracted image for four different expansion factors.  Changes for source-expanded images are shown in the upper panel and those for control images in the lower panel.}
 \label{Plt-Exp_x3_Changes}
\end{figure}

Figs~\ref{Plt-Exp_x3_Example} and \ref{Plt-Exp_x3_Overplots} show results from 1.6 $\mu$m images after expanding source-mask areas by a factor of four. Mean-square fluctuations and associated Gaussian noise in unexpanded, control, and mask-expanded images are plotted in Fig.~\ref{Plt-Exp_x3_Example}. Net fluctuations, calculated as
\begin{equation}
\mathcal{F}^2_\rmn{net}(\theta)= \mathcal{F}^2(\theta)-\mathcal{F}^2_\rmn{noise}(\theta)
\end{equation}
are also shown for each of these images, where symbols are defined in Table~\ref{Tbl-symbols}. When compared to spectra from the unexpanded and control images, the image with source-expanded masks (right panel) has decreased fluctuations, which are most apparent above $\sim$3 arcsec. Direct comparisons of total mean-square fluctuations for these images are shown in Fig.~\ref{Plt-Exp_x3_Overplots}. As shown in the upper panel of Fig.~\ref{Plt-Exp_x3_Changes}, progressive mask expansion causes mean-square fluctuations to noticeably decrease from about 1--50 arcsec relative to the unexpanded image with a maximum reduction of almost 60 per cent in the 10 to 40 arcsec range at an expansion factor of six. While source-mask expansion generally results in a decrease in fluctuations, Figs~\ref{Plt-Exp_x3_Overplots} and~\ref{Plt-Exp_x3_Changes} show an increase at $\sim$80 arcsec where fluctuations in the mask-expanded images exceed those of the unexpanded image. Although the actual difference is small, and on the order of Poisson noise, its effect on a percentage basis is exaggerated by the low fluctuation power at large scales. 

In theory, control images generated by randomly placing individual source-masks on the NUDF background should not alter the underlying power spectrum.  However, the bottom panel of Fig.~\ref{Plt-Exp_x3_Changes} shows that even after correcting for the fraction of masked pixels, the randomly placed source-masks on control images cause distortions in the spectral power which is primarily due to the masking process itself (see Section~\ref{ss-Control}).  Per cent changes become progressively more significant at larger angular scales as the expansion factor is increased and can be as high as almost 40 per cent for an expansion factor of six.
\begin{table*}
 \centering
\begin{minipage}{150mm}
 \caption{Parameter Definitions}
 \label{Tbl-symbols}
 \begin{tabular}{ll}
 \hline
 Abbreviation& Definition\\
 \hline
 $ \mathcal{F}_\rmn{exp}(\theta)$&Fluctuations in the image with expanded source masks \\
 $ \mathcal{F}_\rmn{control}(\theta)$&Fluctuations in the control image. See Section~\ref{ss-Control}\\
 $ \mathcal{F}(\theta)$&Total fluctuations in an image, including noise\\
 $ \mathcal{F}_\rmn{noise}(\theta)$&Fluctuations in a "noise image" with unmasked pixels in original image replaced by Gaussian noise\\
 $ \mathcal{F}_\rmn{net}(\theta)$&Difference between fluctuations in the original image and its associated noise image \\

 \hline
\end{tabular}
All fluctuations are a function of the angular scale, $\theta$\\
\end{minipage}
\end{table*}

\subsection{Contributions of unsubtracted galaxy wings to fluctuation power}\label{ss-results}
\begin{figure}
 \center{\includegraphics[width = \columnwidth,trim = 0in 0in 0in 0in, clip]{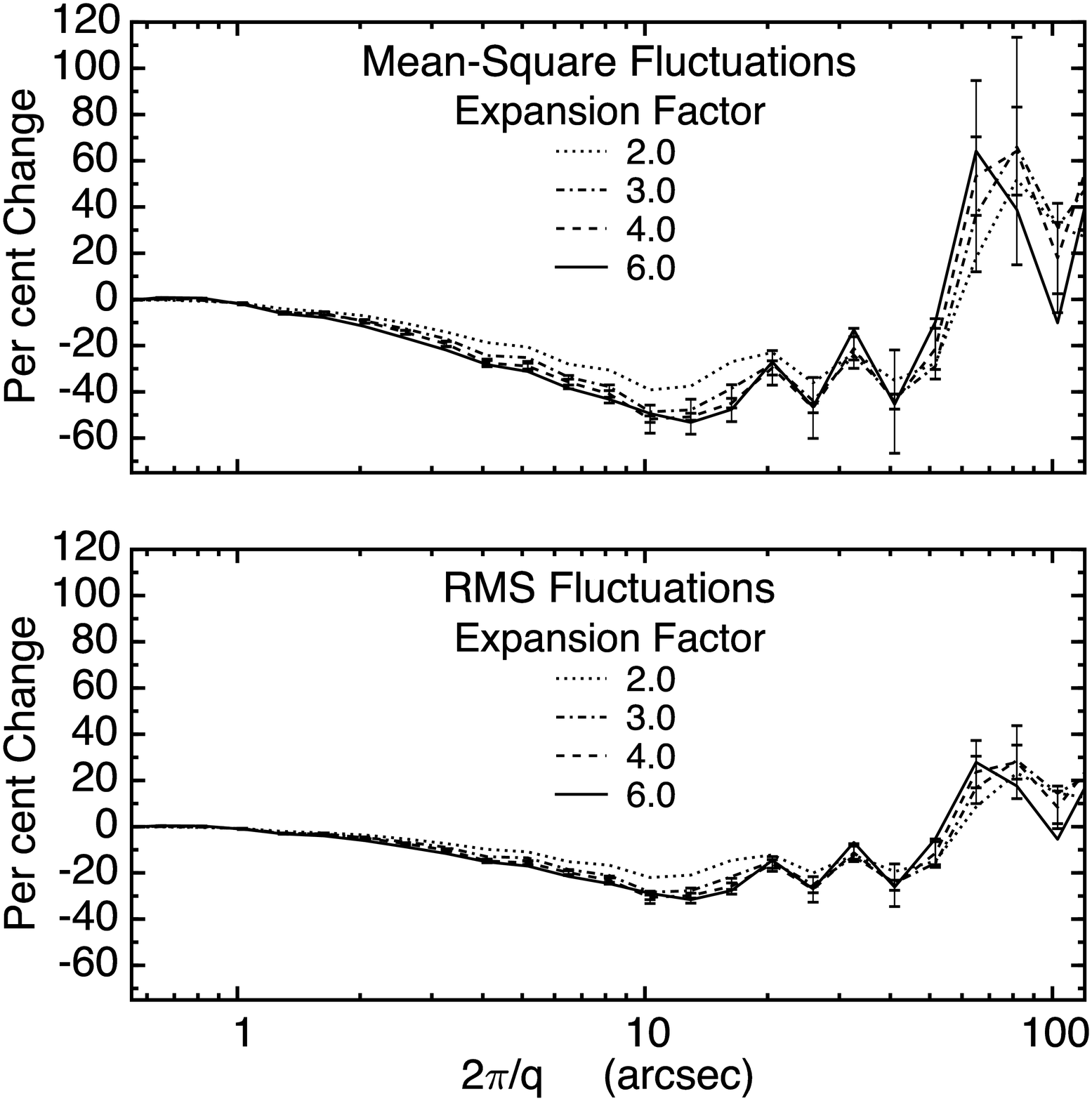}}
 \caption{Mean-square and rms fluctuation changes at 1.6 $\mu$m relative to control images for four different source mask expansion factors. Inner error bars represent one standard deviation for 1000 calculations for an expansion factor of four. Outer error bars also include Poisson errors.}
 \label{Plt-ExpChangesComp}
\end{figure}

Source-mask expansion multiples of 1.1, 1.2, 1.3, 1.4, 1.5, 1.6, 1.7, 1.8, 1.9, 2.0, 2.5, 3.0, 4.0, 5.0 and 6.0 in both filters are tested. At the maximum mask expansion of 6.0, 50 per cent of image pixels are set to zero. To compensate for spectral distortions caused by masking, particularly at high expansion factors, fluctuations in mask-expanded images are compared to the mean of their 1000 associated control images. Fig.~\ref{Plt-ExpChangesComp} shows relative changes after expanding source-mask areas in the 1.6 $\mu$m image. The per cent change in mean-square fluctuation power at angle $\theta,\;\Delta p(\theta)$, is calculated as
\begin{equation}
\Delta p(\theta)=100\times\frac{\mathcal{F}^2_\rmn{exp}(\theta)-\mathcal{F}^2_\rmn{control}(\theta)}{\mathcal{F}^2_\rmn{control}(\theta)},
\end{equation}
and for rms fluctuations as
\begin{equation}
\Delta p(\theta)=100\times\frac{\mathcal{F}_\rmn{exp}(\theta)-\mathcal{F}_\rmn{control}(\theta)}{\mathcal{F}_\rmn{control}(\theta)}.
\end{equation}
Relative changes plateau with increasing source mask expansion and essentially no benefit is seen after an expansion of 4.0. The peak effect at 1.6 $\mu$m is at $\sim$10 arcsec with decreases of 52 $\pm$ 6 per cent for mean-square and 31 $\pm$ 3 per cent for rms fluctuations. The maximum changes at 1.1 $\mu$m are at 40 arcsec with decreases of 50 $\pm$ 25 per cent for mean-square fluctuations and 30 $\pm$ 11 per cent for rms values.

\begin{figure}
 \center{\includegraphics[width = \columnwidth,trim = 0in 0in 0in 0in, clip]{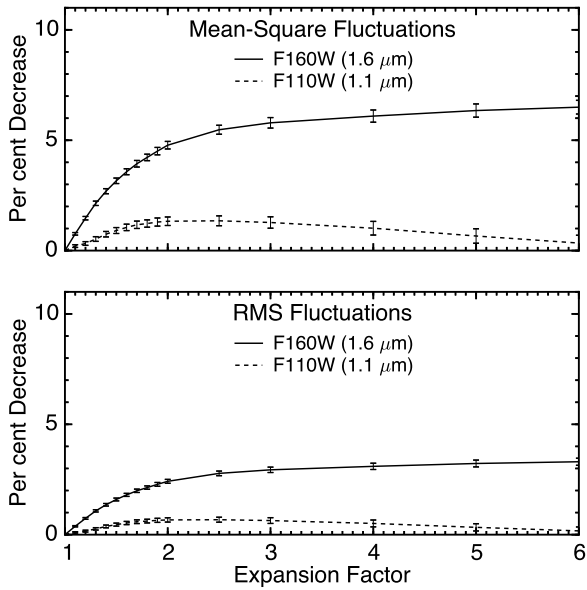}}
 \caption{Integrated fluctuation changes as a function of source mask expansion for F160W and F110W images. Error bars include both one standard deviation for the 1000 calculations and Poisson errors.}
 \label{Plt-SumChanges}
\end{figure}

Fig.~\ref{Plt-SumChanges} shows results for all expansions after summing fluctuations above 0.5 arcsec. The per cent decrease in mean-square fluctuations, $\Delta p_i$, at expansion factor, $i$, is
\begin{equation}
\Delta p_\rmn{i}=100\times\frac{\sum\limits_\rmn{j}\mathcal{F}^2_\rmn{i,\; control}(\theta_\rmn{j})-\sum\limits_\rmn{j}\mathcal{F}^2_\rmn{i,\;exp}(\theta_\rmn{j})}{\sum\limits_\rmn{j}\mathcal{F}^2_\rmn{i, \;control}(\theta_\rmn{j})}
\end{equation}
and for rms fluctuations,
\begin{equation}
\Delta p_\rmn{i}=100\times\frac{\sqrt{\sum\limits_\rmn{j}\mathcal{F}^2_\rmn{i,\; control}(\theta_\rmn{j})}-\sqrt{\sum\limits_\rmn{j}\mathcal{F}^2_\rmn{i,\; exp}(\theta_\rmn{j})}}{\sqrt{\sum\limits_\rmn{j}\mathcal{F}^2_\rmn{i, \;control}(\theta_\rmn{j})}}.
\end{equation}
As suggested in Fig.~\ref{Im-fields}d, subtracting the F110W wings has less effect than subtracting those of the F160W images. No significant benefit is noted in F160W images after individual mask areas are expanded by a factor of $\sim$4.0 with only minor changes above 2.5. Eliminating residual flux in the wings of identified objects reduces total summed mean-square fluctuation power in 1.6 $\mu$m images by only 6.5 $\pm$ 0.3 per cent and rms fluctuations by 3.6 $\pm$ 0.2 per cent. Changes are significantly less at 1.1 $\mu$m with reductions of 1.4 $\pm$ 0.3 and 0.7 $\pm$ 0.1 per cent, respectively.

\begin{figure*}
 \center{\includegraphics[width = \textwidth ,trim = 0in 0in 0in 0in, clip]{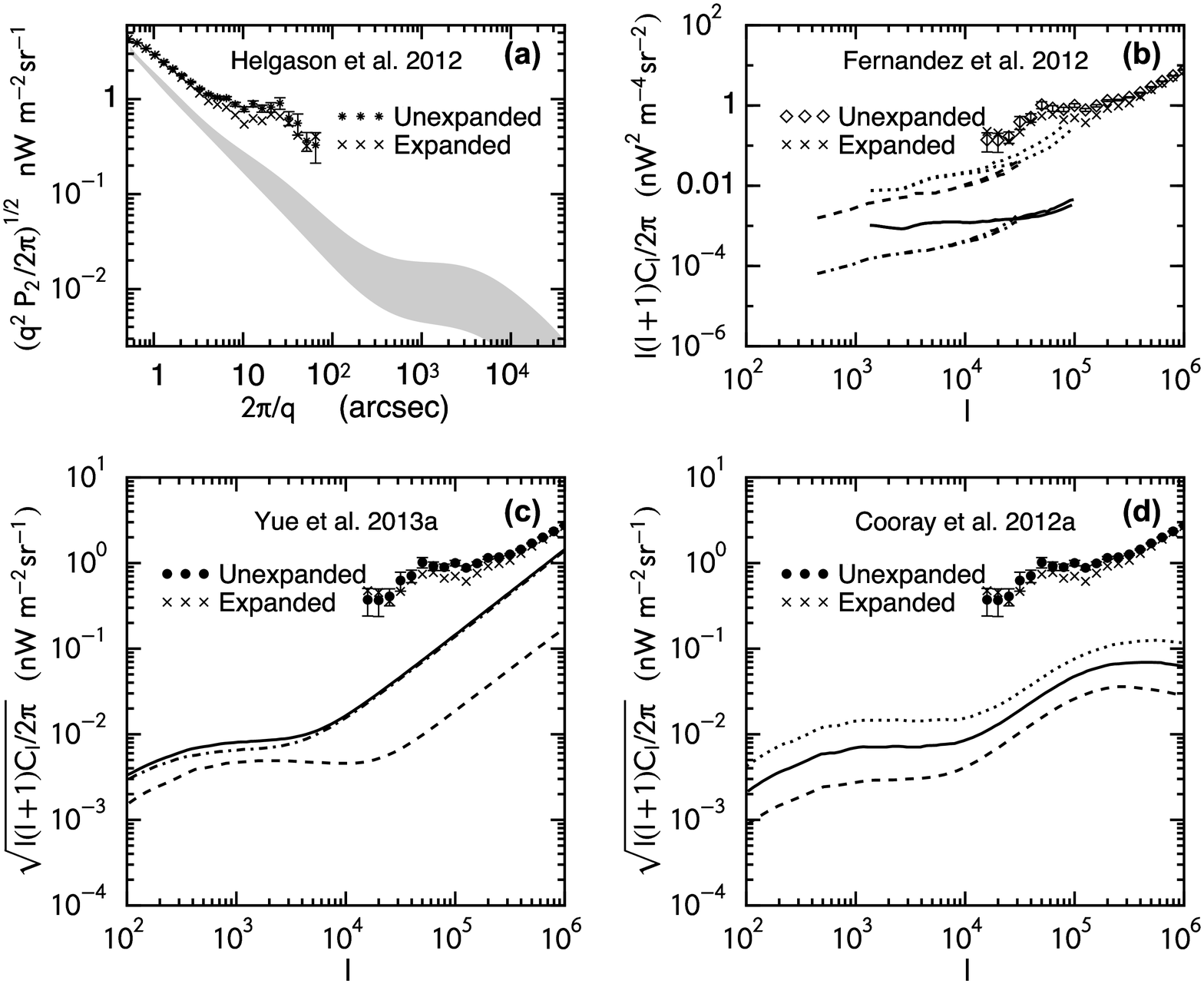}}
 \caption {Fluctuations after subtracting source wings compared to the original (unexpanded) NUDF data for four published theoretical models of the 1.6 $\mu$m NIRB. (a): The shaded region corresponds to the bracketed high (HFE) and low (LFE) faint end areas shown in fig. 9 of \citet{h2}. (b): Simulations from \citet{f1} adapted from their fig. 4, include minimum masses of 10$^8$ M$_\odot$ (solid line) and 10$^9$ M$_\odot$ (dotted line) without suppression as well as models with small haloes suppressed at high (dashed lines) and low (dash-dotted lines) efficiencies. (c): Models shown from fig. 5 of \citet{y1} include contributions to the NIRB of their estimates for z $>$ 5 (dashed lines), \citet{h2} estimates for z $<$ 5 (dash-dotted line), and the calculated total (solid line). (d): The models of halo occupation distributions from fig. 12 of \citet{c3} represent a threshold mass of 10$^6$ M$_\odot$ with efficiency fractions of 0.03, 0.02, and 0.04 and associated ionisation photon escape fractions of 0.5, 0.9, and 0.1. These are shown as the solid, dashed, and dotted lines, respectively. Plot dimension ratios, axis labels, and plot symbols are similar to those in the original publications.  Error bars in the NUDF data represent Poisson noise in the original (unexpanded) images and in many cases are obscured within the plot symbols.}
 \label{Plt-Pubs}
\end{figure*}

As noted above, subtracting residual flux in the 1.6 $\mu$m images reduces mean-square and rms fluctuations by as much as 52 and 31 per cent (Fig.~\ref{Plt-ExpChangesComp}). The significance of these decreases is estimated for publications of four theoretical models of NIRB fluctuations in which the authors have cited NUDF results to support their conclusions \citep{h2,f1,y1,c3}. Fig.~\ref{Plt-Pubs} compares fluctuations at 1.6 $\mu$m after a source-mask expansion of 4.0 to the original NUDF data used in these publications. Data points from cited plots were extracted using online digitising software.\footnote{http://arohatgi.info/WebPlotDigitizer} While an expansion of 4.0 effectively eliminates unsubtracted flux in galaxy wings (Fig.~\ref{Plt-ExpChangesComp}), 38 per cent of total pixels are set to zero. As discussed in Section~\ref{ss-Effects}, and shown in Figs~\ref{Plt-Exp_x3_Changes} and~\ref{Plt-ExpChangesComp}, if spectral distortions caused by the masking process itself are ignored, changes will be overestimated at angular scales larger than $\sim$10 arcsec.  Comparing fluctuation spectra of source-mask expanded images to control images with the same number of subtracted pixels provides a better compensation for masking effects than only correcting for source-subtracted areas. Therefore, mean-square and rms fluctuations for the mask-expanded images are calculated by multiplying their original, unexpanded values by ${\mathcal{F}^2_\rmn{exp}(\theta)}/{\mathcal{F}^2_\rmn{control}(\theta)}$ and ${\mathcal{F}_\rmn{exp}(\theta)}/{\mathcal{F}_\rmn{control}(\theta)}$, respectively. \textit{As shown in Fig.~\ref{Plt-Pubs}, subtracting residual flux from source wings would not affect any published conclusions drawn from the previously reported NUDF fluctuations.} 

 \section{Discussion and Conclusions}
This study has shown that faint wings of source-subtracted galaxies account for only a small portion of total NUDF background fluctuation power above 0.5 arcsec. These results are generally consistent with estimates of the contributions of residual galaxy wings to spectral power in \textit{Spitzer}/IRAC deep images \citep{k3, a1}. These prior studies did not identify individual sources and galaxy wings were removed by equally increasing the size of all masked regions \citep{a1}. Radial mask expansion was limited to two pixels in order to retain enough unmasked pixels for reliable Fourier analysis. The higher resolution of the NUDF observations used in the present study allows residual flux to be eliminated using an algorithm that expands individual source masks proportional to the size of the underlying galaxy.  This reduces the number of expanded pixels and allows a more robust analysis of the effects of galaxy wings on measured fluctuations. \citet{k3} found a decrease of less than a few per cent in $P_2(q)$ after radially expanding masks by two pixels while \citet{a1} noted that in most cases the largest changes in rms fluctuations were less than the 1$\sigma$ uncertainties. Both of these findings are significantly less than the decreases observed in this study (Figs~\ref{Plt-ExpChangesComp} and~\ref{Plt-Pubs}) and probably result from a more complete source-subtraction in their original images.

A limitation to the methodology used in this study is the degree to which control images can compensate for spectral distortions caused by source-mask expansion. Individual source-masks in control images have a smaller average size and a more even distribution than those in the source-mask expanded images and, therefore, will affect underlying fluctuations differently. Although using control masks as generated in this study will not fully correct for spectral distortions caused by mask expansion, they do provide a better compensation than only dividing by the fraction of unmasked pixels. This is important because, as shown in Figs~\ref{Plt-Exp_x3_Changes} (upper panel) and~\ref{Plt-ExpChangesComp}, both the magnitude of changes and the expansion factor required to eliminate wings would be overestimated if control images were not used.

The IHL model for fluctuations applied to observations from both the \textit{Spitzer} Deep, Wide-Field Survey (Cooray et al. 2012b) and CIBER (Zemcov et al. 2014) predicts effects becoming significant at angular scales larger than those encompassing the NUDF field. Yue et al. (2013b) fit their DCBH model for fluctuations at 3.6 and 4.5 $\mu$m to scales $>$ 100 arcsec and, as suggested in fig. 4 of their paper, DCBHs would have negligible contributions at 1.6 and 1.1 $\mu$m. Therefore, studies postulating IHL or DCBHs to account for the NIRB did not use NUDF results to support their hypotheses.

As shown in Fig.~\ref{Plt-Pubs}, even the maximum decreases for mean-square and rms fluctuations are barely noticeable on a log scale. Results after eliminating source wings from NUDF images represent the upper limits of fluctuations at 1.1 and 1.6 $\mu$m and continue to remain well above those predicted by models for known galaxy populations \citep{h2} or high-redshift galaxies \citep{f1, y1, c3}. The revised fluctuation spectra presented in this paper would not affect any of these conclusions. These results exclude previous suggestions that extended wings of detected galaxies may be a major contributor to fluctuations in source-subtracted NUDF images and confirm that other explanations are required to account for most of the NIRB fluctuations at the angular scales of NUDF observations.

\section*{Acknowledgments}

I would like to thank Professor Rodger Thompson for his patient guidance, discussions and encouragement.  I also want to acknowledge the very helpful review and suggestions made by the anonymous referee.

\bsp
\label{lastpage}

\begin{thebibliography}{99}

\bibitem[\protect\citeauthoryear{Arendt et al.}{2010}]{a1} Arendt R.~G., Kashlinsky A., Moseley S.~H., Mather J., 2010, ApJS, 186, 10 

\bibitem[\protect\citeauthoryear{Bertin \& Arnouts}{1996}]{b1} Bertin E., Arnouts S., 1996, A\&AS, 117, 393 

\bibitem[\protect\citeauthoryear{Cooray et al.}{2007}]{c2} Cooray A., et al., 2007, ApJ, 659, L91

\bibitem[\protect\citeauthoryear{Cooray et al.}{2012a}]{c3} Cooray A., Gong Y., Smidt J., Santos M.~G., 2012a, ApJ, 756, 92 

\bibitem[\protect\citeauthoryear{Cooray et al.}{2012b}]{c4} Cooray A., et al., 2012b, Natur, 490, 514 

\bibitem[\protect\citeauthoryear{Fernandez et al.}{2012}]{f1} Fernandez E.~R., Iliev I.~T., Komatsu E., Shapiro P.~R., 2012, ApJ, 750, 20 

\bibitem[\protect\citeauthoryear{Helgason, Ricotti, \& Kashlinsky}{2012}]{h2} Helgason K., Ricotti M., Kashlinsky A., 2012, ApJ, 752, 113 

\bibitem[\protect\citeauthoryear{Kashlinsky et al.}{2005}]{k3} Kashlinsky A., Arendt R.~G., Mather J., Moseley S.~H., 2005, Natur, 438, 45 

\bibitem[\protect\citeauthoryear{Kashlinsky et al.}{2012}]{k5} Kashlinsky A., Arendt R.~G., Ashby M.~L.~N., Fazio G.~G., Mather J., Moseley S.~H., 2012, ApJ, 753, 63 

\bibitem[\protect\citeauthoryear{Matsumoto et al.}{2011}]{m3} Matsumoto T., et al., 2011, ApJ, 742, 124

\bibitem[\protect\citeauthoryear{Szalay, Connolly, \& Szokoly}{1999}]{s1} Szalay A.~S., Connolly A.~J., Szokoly G.~P., 1999, AJ, 117, 68 

\bibitem[\protect\citeauthoryear{Thompson et al.}{2005}]{t3} Thompson R.~I., et al., 2005, AJ, 130, 1 

\bibitem[\protect\citeauthoryear{Thompson et al.}{2006}]{t2} Thompson R.~I., Eisenstein D., Fan X., Dickinson M., Illingworth G., Kennicutt R.~C., Jr., 2006, ApJ, 647, 787 

\bibitem[\protect\citeauthoryear{Thompson et al.}{2007a}]{t1} Thompson R.~I., Eisenstein D., Fan X., Rieke M., Kennicutt R.~C., 2007a, ApJ, 657, 669 

\bibitem[\protect\citeauthoryear{Thompson et al.}{2007b}]{t4} Thompson R.~I., Eisenstein D., Fan X., Rieke M., Kennicutt R.~C., 2007b, ApJ, 666, 658

\bibitem[\protect\citeauthoryear{Yue et al.}{2013a}]{y1} Yue B., Ferrara A., Salvaterra R., Chen X., 2013a, MNRAS, 431, 383 

\bibitem[\protect\citeauthoryear{Yue et al.}{2013b}]{y2} Yue B., Ferrara A., Salvaterra R., Xu Y., Chen X., 2013b, MNRAS, 433, 1556 

\bibitem[\protect\citeauthoryear{Zemcov et al.}{2014}]{z1} Zemcov M., et al., 2014, Sci, 346, 732 

\end{thebibliography}
\end{document}